\documentclass[twocolumn,twoside,slac_two]{revtex4}
\usepackage{graphicx}
\usepackage{fancyhdr}
\begin{document}

\title{Population study for  $\gamma$-ray emitting 
 Millisecond Pulsars and $Fermi$ unidentified sources }

%

\author{J.Takata, Y.Wang, K.S. Cheng}
\affiliation{Department of Physics, University of Hong Kong, 
Pokfulam Road, Hong Kong}

\begin{abstract}
The $Fermi$-LAT has revealed that rotation powered millisecond pulsars 
(MSPs) are a major contributor to the Galactic $\gamma$-ray source population. 
We discuss the $\gamma$-ray emission process  within 
the context  of the outer gap accelerator model, and use a Monte-Calro method 
to simulate the Galactic population of the $\gamma$-ray emitting MSPs.  
We find that the outer gap accelerator controlled by the
 magnetic pair-creation process is preferable in explaining 
the possible correlation between the $\gamma$-ray luminosity and 
 the spin down power. Our Monte-Calro simulation implies that most of 
the $\gamma$-ray emitting MSPs are radio quiet in the present sensitivity of 
the radio survey, indicating that most of the $\gamma$-ray
MSPs have been unidentified.  We argue that the Galactic $Fermi$ 
unidentified sources located at  high latitudes should
be dominated by MSPs, whereas the sources in the galactic plane
are dominated by radio-quiet canonical pulsars.
\end{abstract}

\maketitle

\thispagestyle{fancy}


\section{Introduction}
The $Fermi$ Large Area Telescope ($Fermi$-LAT) has discovered
 over  80 $\gamma$-ray pulsars, and revealed that the $\gamma$-ray 
pulsars are a major class of Galactic $\gamma$-ray sources. Of 
these, $Fermi$-LAT first detected pulsed $\gamma$-ray emission 
from 11 millisecond pulsars (Abdo et al. 2010a,
 2009a,b; Saz Parkinson et al. 2010; Guillemot et al. 2011).   
Furthermore, the detection of  radio millisecond pulsars (MSPs) 
 associated with  over  20 unidentified $Fermi$ point sources 
(e.g. Ransom et al. 2011; Keith et al. 2011) 
has been reported, suggesting  that the millisecond pulsar, 
 as well as the canonical pulsar is  one of the 
major Galactic $\gamma$-ray source.   
The $\gamma$-ray emission from pulsars have been discussed in 
the context of a polar cap accelerator 
(Ruderman \& Sutherland  1975), a slot gap  
(Harding \& Muslimov 2011) and an outer gap accelerator (Cheng, Ho \& Ruderman 1986; Takata, Wang \& Cheng 2010b).  

The high quality data measured  by the $Fermi$ enable us to 
perform a  detail study for   population 
of the $\gamma$-ray pulsars.  For example, Takata et al.  (2010b) and 
Takata, Wang and Cheng (2011d) have studied 
the possible relation between the high-energy emission properties 
and pulsar spin down power in the context of the outer gap accelerator model.
 They proposed  that the outer gap accelerator model controlled by 
the magnetic pair-creation process  can explain  the observed population 
statistics better than that  controlled by  the photon-photon 
pair-creation process (e.g. Zhang \& Cheng 2003). 
Story, Gonthier \& Harding (2007) studied the 
population of $\gamma$-ray millisecond pulsars within the context of 
the slot gap accelerator model, and predicted the $Fermi$ observations.
 They predicted that the $Fermi$ will detect 12 radio-loud and 
33-40 radio-quiet $\gamma$-ray millisecond pulsars.  
With the Monte-Calro simulation of the outer gap,  
Takata, Wang \&  Cheng (2011a,b) have explained 
the observed  distributions of the characteristics 
of the  $\gamma$-ray pulsars detected by the $Fermi$ with the 
six-month long  observation. 
The  population studies  (e.g.	Kaaret \&  Philip 1996;
 Takata et al. 2011b,c) have also pointed out that unidentified 
 MSPs located at high-Galactic latitudes 
will associate with the $Fermi$ unidentified sources (Abdo et al. 2010b). 

In this proceeding, we will review our recent Monte-Carlo studies 
for the Galactic population of the 
$\gamma$-ray emitting MSPs (Takata et al. 2010b; Takata et al. 2011a,b,c)
 and the possible association with the $Fermi$ unidentified sources. 

\section{$\gamma$-ray emission from outer gap}
\label{gamma}

\subsection{$\gamma$-ray luminosity}
For the outer gap model, the luminosity of the $\gamma$-ray emissions 
is typically 
\begin{equation}
L_{\gamma}\sim f^3L_{sd},
\label{lgamma}
\end{equation}
where $L_{sd}$ is the spin down power of the pulsar and the gap fraction 
$f$ is 
defined as the ratio of the gap thickness at the light cylinder 
to the light cylinder radius $R_{lc}=Pc/2\pi$, 
where $P$ is pulsar rotation period.

Zhang \& Cheng (2003) have argued  a self-consistent outer gap model 
controlled by  the photon-photon pair-creation process between 
the curvature photons and the X-rays from the stellar surface. 
 They estimated the gap fraction for the MSPs  as 
\begin{equation}
f_{p}=7.0\times 10^{-2}(P/1~\mathrm{ms})^{26/21}(B_s/10^8\mathrm{G})^{-4/7}
\delta r_{5}^{2/7},
\label{fp}
\end{equation}
where $B_s$ is the stellar magnetic field of the global dipole field and 
  $\delta r_5$ is the distance (in units of $10^{5}$~cm)
 from the stellar surface to the position, where the local multiple 
 magnetic field, which dominates the global dipole field, 
 is comparable to the dipole field, and it will be $\delta r_5\sim 1-10$~cm.

 Takata et al. (2010b) argued that the incoming particles emit photons 
with an energy $m_ec^2/\alpha_f\sim 70 \rm MeV$ by curvature radiation 
near the stellar surface and that  these 
photons can become pairs via the magnetic 
pair creation process.  For a simple dipole field 
structure, all pairs move inward and cannot affect the outer gap accelerator.  
However  if the local field lines near the surface are bent sideward due
 to the strong multipole field, the pairs created in these local magnetic 
field lines can have an angle greater than 90$^{\circ}$, which results in 
an outgoing flow of pairs.  In this model, the fractional gap 
thickness in this circumstance is
\begin{equation}
f_m\sim0.025KP^{1/2}_{-3},
\label{fm}
\end{equation}
where $K\sim B_{m,12}^{-2}s_7$ characterizes the local parameters. 
Here, $B_{m,12}$ and $s_7$ are the strength of the 
local  magnetic field in units
 of $10^{12}$G and the local curvature radius in units of $10^7$cm, 
respectively, near the stellar surface. 

The $\gamma$-ray luminosity $L_{\gamma}$ (\ref{lgamma})  can be cast in 
terms of the spin down power $L_{sd}= 2(2\pi)^4 \mu /(3c^3P^4)$ as  
\begin{equation}
L^p_{\gamma}\sim 10^{32}L_{sd,34}^{1/14}B_{8}^{1/7}
\delta_{5}^{6/7}
~\mathrm{erg~s^{-1}},
\label{lglsp}
\end{equation}
 for the outer gap controlled by the photon-photon pair-creation process, and 
 \begin{equation}
L^m_{\gamma}\sim 6\times 10^{32}L_{sd,34}^{5/8}K_1^{3}B_{8}^{3/4}
~\mathrm{erg~s^{-1}},
\label{lglsm}
\end{equation}
by the magnetic pair-creation process.  Here, $L_{sd, 34}=(L_{sd}/10^{34}~
\mathrm{erg~s^{-1}})$, $K_1=K/10$ and $B_8=B_s/10^8~\mathrm{G}$ .

\begin{figure}
 \includegraphics[width=8cm, height=8cm]{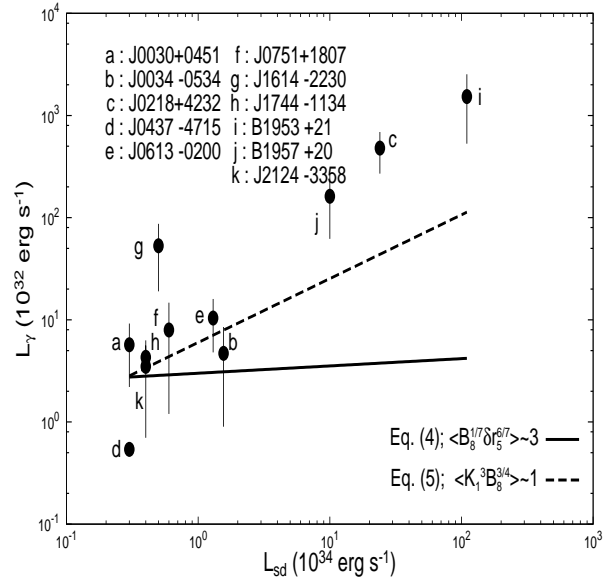}
 \caption{The $\gamma$-ray luminosity and the spin down power for the MSPs. 
The sole and dashed line are  the relation predicted by the outer gap 
model controlled by the photon-photon pair-creation process and by 
the magnetic pair-creation process, respectively. The observed luminosity is  
$L^{ob}_{\gamma}=4\pi d^2F$ with $d$ being the distance 
and $F$ the observed flux.  See Takata et al. 2011d.}
\label{lgls}
 \end{figure}

In Figures~\ref{lgls}  the model predictions given by equations~(\ref{lglsp})
 and (\ref{lglsm}) are plotted  with the solid line 
 or dashed line, respectively. The filled
 circles represent the MSPs detected by the $Fermi$-LAT.  
 Notwithstanding 
the large observational errors, the data points at large $L_{\gamma}$ 
in Figures~\ref{lgls}  may suggest that the magnetic pair-creation model 
for the gap closing  process is  preferred over the photon-photon 
pair-creation model for the $L_{\gamma} - L_{sd}$ relation.

\begin{figure}
 \includegraphics[width=8cm, height=8cm]{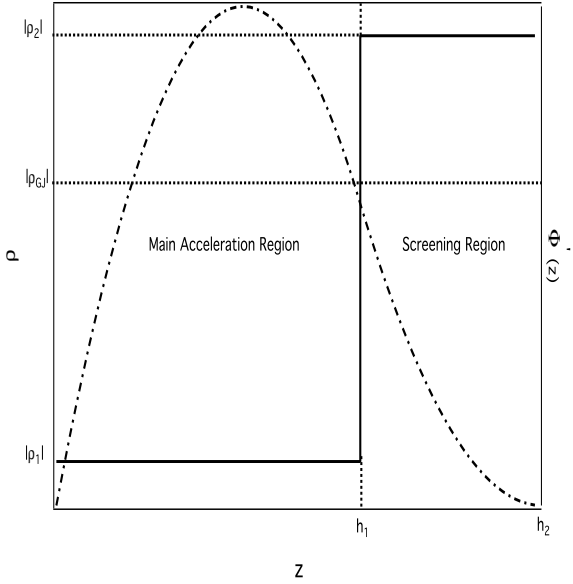}
 \caption{The simplified distribution of the charge density (solid line) and
the corresponding accelerating potential (dot–dashed line) of the two-layer
outer gap. See Wang et al. 2011. }
\label{two}
 \end{figure}
\subsection{Two layer outer gap model}
A simple description of the electric field structure inside 
 the two-layer outer  gap is discussed in  Wang, Takata \& Cheng (2010). 
 We illustrate the picture for the gap 
structure in Figure~\ref{two}, where  we denote 
 $z$  as the coordinate perpendicular to the magnetic 
field line in the poloidal plane, and $\rho$ is the charge density.  
 In the main acceleration region, the charge density is $\sim$10\% 
of the Goldreich-Julian value, and 
 an electric field along the magnetic field line 
 accelerates the electrons and positrons up to 
the Lorentz factor of $\Gamma\sim 10^{7.5}$ and the accelerated 
particles emit $\sim$ GeV photons via the curvature radiation process. 
In the screening region,  the large number of pairs created by 
the pair-creation process starts to screen out  the gap electric field. 

\subsubsection{Charge density distribution}
The previous electrodynamics studies expect that  the particle number 
density   increases  exponentially  near the  boundary ($z=h_1$) 
between the main acceleration and screening regions (Cheng et al. 1986), 
we approximately describe  the distribution of the charge density  in  the $z$-direction  
 as follows  (Fig.~\ref{two})
\begin{equation}
\rho(\vec{r})=\left\{
   \begin{array}{ccc}
            \rho_1(x,\phi), &$if$& 0\leq z\leq h_1(x,\phi),\\
            \rho_2(x,\phi), &$if$& h_1(x,\phi)<z\leq h_2(x,\phi),
   \end{array}\right.
\end{equation}
where  $x$ and $\phi$ represent the coordinates along the magnetic 
field and the azimuthal direction.  In addition,  
$z=0$ and $z=h_2$ correspond to  the last-open field line and 
the upper boundary of the gap, respectively. For simplicity,  
we  assume that  the upper boundary ($h_2$) as well as  
the boundary  ($h_1$)  is defined with a given magnetic field 
line. Because the charge density in the screening region will 
 be proportional to the Goldreich-Julian charge density, 
 we approximate that $\rho-\rho_{GJ}\sim g(z,\phi)\rho_{GJ}(\vec{r})$, where 
\begin{equation}
g(z,\phi)=\left\{
   \begin{array}{ccc}
            -g_1(\phi), &$if$& 0\leq z\leq h_1(x,\phi),\\
            g_2(\phi), &$if$& h_1(x,\phi)<z\leq h_2(x,\phi).
   \end{array}\right.
\end{equation}
and  $g_1>0$ and $g_2>0$ so that $|\rho|<|\rho_{GJ}|$ 
for the main  acceleration  region and $|\rho|>|\rho_{GJ}|$ for 
the screening region. 
In this paper, by neglecting the $z$-dependence of the Goldreich-Julian charge 
density, we approximate as $\rho_{GJ}(x,\phi)\sim -\Omega B x/2\pi c R_c$, 
where $\Omega$ and $R_c$ are the angular frequency of the pulsar 
 and the curvature radius of the field line, respectively.

\subsubsection{Accelerating electric field}
\label{electric}
To  obtain the  typical strength of the electric field in the gap,  
we find the solution of  the Poisson equation 
for each azimuthal angle (Wang et al. 2010),
\begin{equation}
 \frac{\partial^2{}}{\partial{z^2}}\Phi'(x,z,\phi)|_{\phi=\mathrm{fixed}}
=-4\pi [\rho(x,z,\phi)-\rho_{GJ}(\vec{r})]_{\phi=\mathrm{fixed}},
\label{Poisson1}
\end{equation}
where $\Phi'$ is the  electric potential of the accelerating field.  
The boundary conditions on the lower ($z=0$) and upper ($z=h_2$) boundaries 
are given by  $\Phi'(x,z=0,\phi)=\Phi'(x,z=h_2,\phi)=0$.  
Imposing $\Phi'$ and $\partial\Phi'/\partial z$ are continuous 
at the boundary  $z=h_1$, we eventually obtain the solution 
of the accelerating electric field ($E_{||}=-\partial \Phi'/\partial x$) as  
\begin{equation}
E_{||}(\vec{r})\sim\frac{\Omega Bh^2_2(x,\phi)}{cR_s}[
-g_1(\phi)z'^2+C_1(\vec{r})z'],
\label{electric1}
\end{equation}
for $0\le z'\le h_1/h_2$ and 
\begin{equation}
E_{||}(\vec{r})\sim\frac{\Omega Bh^2_2(x,\phi)}{cR_s}
[g_2(\phi)(z'^2-1)+D_1(\vec{r})(z'-1)]
\label{electric2}
\end{equation}
for $h_1/h_2<z'\le 1$, where $z'=z/h_2$, 

\[
 C_1(x,\phi)=-
[g_1
h_1(h_1-2h_2)
+g_2(h_1-h_2)^2]/h_2^2,
\]
and 
\[
 D_2(x,\phi)=-(g_1h_1^2
+g_2h_2^2)/h_2^2. 
\]
In addition,  we used the relations of the dipole field 
that  $\partial(Bh^2_2)/\partial x\sim 0$, 
$\partial z'/\partial x\sim 0$, 
$\partial(h_1/h_2)/\partial x\sim 0$, and approximated that 
$\partial R_c/\partial x\sim 0$. 

On the upper boundary, the total potential field   
(co-rotational + non co-rotational potentials)  in the 
gap is continuously connected
to the co-rotational potential field outside the gap, giving 
$\partial \Phi'/\partial z|_{z=h_2}=-E_{\perp}(z=h_2)=0$, 
yielding the relation between $(h_1,~h_2)$ and $(g_1,~g_2)$ as 
\begin{equation}
(h_2/h_1)^2=1+(g_1/g_2).
\label{condition}
\end{equation}
The gap thickness $h_2$ is calculated that 
$h_2\sim fR_{lc}$,   where $f$ is the fractional gap thickness given by  
Eq.(\ref{fp}) or Eq. (\ref{fm}). In this paper,  we present the results 
using  the ratio of $h_1/h_2=0.95$  and 
the dimensionless charge density of  main acceleration region 
$1-g_1=0.3$~(Takata et al. 2011c). 
The normalized charge density of the screening region, 
$g_2$ is calculated from the equation~(\ref{condition}). 
The Lorentz factor of the accelerated particles is obtained by 
the force balance between the electric field and the curvature radiation drag 
force as 
\begin{equation}
\Gamma=\left(\frac{3R_c^2}{2e}E_{||}\right)^{1/4}.
\end{equation}

\begin{figure}
 \includegraphics[width=8cm, height=8cm]{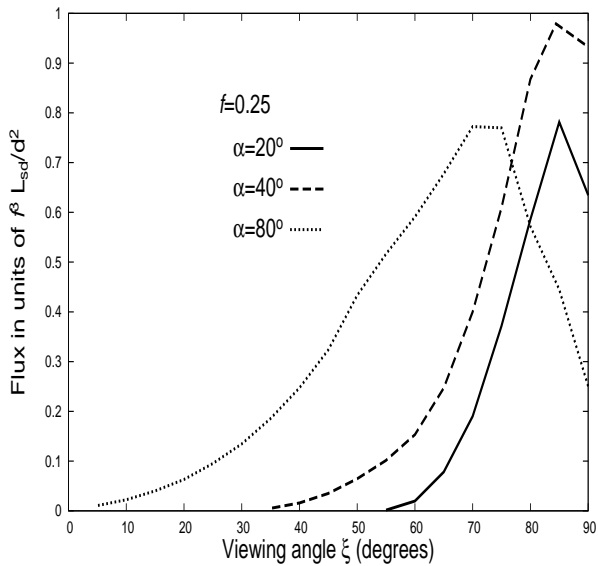}
 \caption{Dependency of the observed $\gamma$-ray flux on the 
viewing angle ($\xi$) and the inclination angle ($\alpha$). $f=f_p=0.25$. 
See Takata et al. (2011c).}
\label{fluxMS}
 \end{figure}

To examine  the dependency of $\gamma$-ray emission properties 
 on the viewing geometry, we explore the typical  3-D geometry of 
the outer gap (Takata et al. 2011c). We calculate the curvature 
radiation process for each emitting point,
 at which  the curvature photons are emitted in the direction of the 
particle motion.  Figure~\ref{fluxMS} shows the  
$\gamma$-ray flux ($\ge 100$~MeV) as a function of 
 the viewing angle $\zeta$ and of 
the inclination  angle $\alpha$.  Here we calculated 
 the rotation period from $B_s=3\times 10^{12}~$G and 
the gap fraction $f=f_{p}=0.25$.  We can see  that 
the calculated flux tends to decrease as the line  of sight  approaches  to 
the rotation axis, where $\zeta=0^{\circ}$.  This is because a large viewing 
angle ($\zeta \rightarrow 90^{\circ})$ can encounter more 
intense emission region, whereas the small viewing angle ($\zeta \ll 
90^{\circ}$) will encounter the less intense region 
 or even miss  the emission region.  Also we find in Figure~\ref{fluxMS}, 
a larger inclination angle shows a larger observed flux for a smaller 
Earth viewing angle. These dependences  
of the $\gamma$-ray flux on the viewing geometry  implies
 that  the $Fermi$ has likely detected a greater number of the 
 pulsars with larger inclination angles  and  larger viewing angles near $90^{\circ}$.  

\section{Monte-Carlo simulation}
\label{monte}
We simulate that Galactic distribution of MSPs using the Monte-Carlo method 
developed by the previous 
studies (e.g. Sturner \& Dermer 1996). We would like to remark that for MSPs, 
the Galactic distribution will  not depend 
on the spin down age of the MSPs.  With the typical velocity of the
observed MSPs ,  $V\sim 70$~km/s,  it is expected that the 
displacement of the MSPs (or binary system) 
with the typical  age, $\ge 100$~Myr, 
becomes larger than the size of the Galaxy. With the  relatively 
 slow velocity, $V\le 100$~km/s,  however, 
the MSPs remain bound in the Galactic potential  and hence
their Galactic distribution does  not  depend on the spin down age.  
Because the independence of the Galactic distribution  on 
the spin down age, we assign the ``current'' pulsar properties  for 
simulated MSPs; namely, 
 we (1) randomly select the age of the simulated
MSP  up to $10$~Gyr,  (2) shifts the simulated MSP from its birth location 
 to the current location,   and (3) assign the  parameters of the MSP 
following the observed  distributions.

For each simulated MSPs, we assume  the radio cone is centered 
on the magnetic axis with a width described by
 the model of Kramer \&  Xilouris (2000). Because 
 the $\gamma$-ray flux depends on the viewing geometry (Figure~\ref{fluxMS}), 
we assume that  the viewing angle and the inclination angle are 
randomly distributed for each simulated MSP.

For the sensitivity of the $Fermi$ observations, we refer the sensitivity 
of $Fermi$  the six-month long observations (Abdo et al. 2010b) 
for the radio selected MSPs.  However, 
there are no  detections of the $\gamma$-ray-selected MSPs
 so far, we cannot simulate the $Fermi$
 sensitivity of the blind search of the MSPs. In this paper, therefore,
 we simulate the population of the $\gamma$-ray-selected MSPs with   
 the $Fermi$ sensitivity of the blind search of canonical pulsar. To simulate
  a longer $Fermi$ observation, we scale the sensitivity as  
$\propto \sqrt{T}$, where $T$ is the length of the observation time.

\section{Results}
We present the results of the Monte-Carlo simulation 
within context of the outer gap model controlled by the magnetic pair-creation process, that is, the fractional gap thickness is given by equation~\ref{fm}.

\begin{figure}
 \includegraphics[width=8cm, height=8cm]{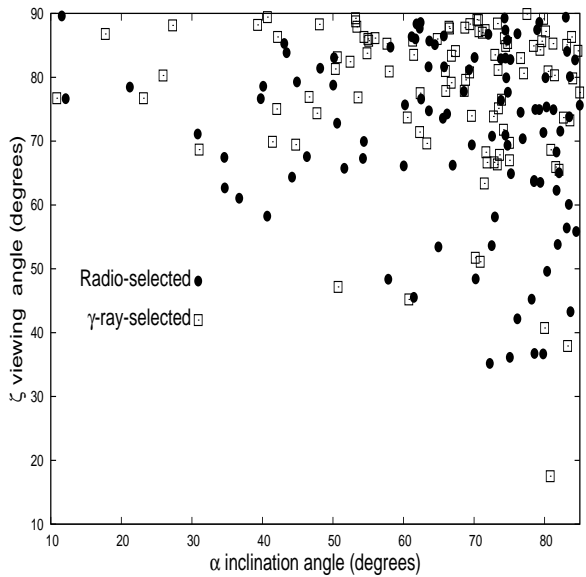}
 \caption{The inclination  axis $\alpha$ and the viewing angle $\zeta$ for 
the simulated  $\gamma$-ray MSPs. Filled circles: Radio selected 
pulsars. Boxes: $\gamma$-ray-selected pulsars. See Takata et al. (2011c).}
\label{axiMS}
 \end{figure}
\begin{figure}
 \includegraphics[width=8cm, height=8cm]{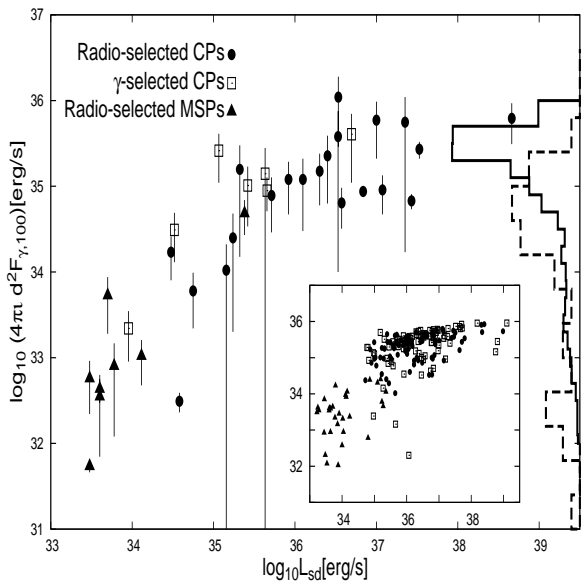}
 \caption{The $\gamma$-ray luminosity  versus the spin down power 
for $\gamma$-ray emitting MSPs. 
The 200 samples of the simulated pulsars are plotted  in the sub-figure. 
See Takata et al. (2011c).}
\label{fluxdis}
 \end{figure}

\subsection{$L_{\gamma}$ vs. $L_{sd}$}
Figure~\ref{axiMS} plots the distribution of  the 
inclination angle ($\alpha$) and the viewing angle ($\zeta$)
 for the simulated  $\gamma$-ray MSPs. 
The filled-circles and the boxes represent 
the radio-selected and $\gamma$-ray-selected $\gamma$-ray
MSPs, respectively.  As we can seen in Figure~\ref{axiMS}, 
no $\gamma$-ray pulsars  are detected with a  smaller inclination
 angles   and a smaller viewing angles. 
This is because the $\gamma$-ray flux decreases 
as the viewing angle and/or  the inclination angle 
decrease (c.f. Figure~\ref{fluxMS}). Hence, 
 our simulation result predicts that a greater number of  MSPs with 
larger inclination  and larger  viewing angles 
($\sim 90^{\circ}$) have 
been likely  detected by the $Fermi$  observations.

Figures~\ref{fluxdis} shows the relation between the 
$\gamma$-ray luminosity and the spin down power. In  the figure,  
we plot the $Fermi$ data with errors taken from Abdo et al. (2010a) 
and Saz~~Parkinson et al. (2010), and present the simulated 200 pulsars 
 in the sub-figures (except for the $\gamma$-ray-selected MSPs). The solid  and
  dashed histograms represent the distributions
 for the simulated and observed $\gamma$-ray pulsars, respectively.
 The present model predicts 
that most  $\gamma$-ray canonical pulsars  have a spin down power of 
 $L_{sd}\sim 10^{35-38}~\mathrm{erg/s}$ and a $\gamma$-ray luminosity  
of  $L_{\gamma}\sim 10^{34-36}$, while  MSPs have  a 
 $L_{sd}\sim 10^{33-35}~\mathrm{erg/s}$ and
 $L_{\gamma}\sim 10^{32.5-34.5}~\mathrm{erg/s}$, which are consistent with 
the observations.  

In Figure~\ref{fluxdis}, we may see that  the spin down power  $L_{sd}$ and 
the $\gamma$-ray luminosity $L_{\gamma}$ of the simulated 
pulsars can be related as $L_{\gamma}\propto 
L_{sd}^{\beta}$ with $\beta\sim 0$ for $L_{sd}\ge 10^{35-36}~\mathrm{erg/s}$ 
and $\beta\sim 0.5$ for $L_{sd}\le 10^{35-36}~\mathrm{erg/s}$.  Because 
  the $\gamma$-ray luminosity is 
proportional to $L_{\gamma}\propto f^3L_{sd}$ [Eq.~(\ref{lgamma})],
 the change of the slope implies the  switching gap closure process
  between the photon-photon pair-creation process, which predicts 
$L_{\gamma}\propto   L_{sd}^{1/14}$ [Eq.~(\ref{lglsp})] and the 
magnetic pair-creation process, which predicts  
$L_{\gamma}\propto L_{sd}^{5/8}$  [Eq.~(\ref{lglsm})].  
\begin{table}
\begin{tabular}{ccccccc}
\hline
 & \multicolumn{2}{c}{six-months} & \multicolumn{2}{c}{five-years} & \multicolumn{2}{c}
{ten-years} \\ MSPs
 & $N_r$ & $N_g$ & $N_r$ & $N_g$ & $N_r$ & $N_g$ \\
\hline\hline
Ra. Sen. (x 1) & 10 & 52 & 14 & 200 & 16 & 284 \\
Ra. Sen. (x 2) & 16 & 48 & 26 & 190 & 29 & 274 \\
Ra. Sen. (x 10) & 45 & 32 & 82 & 152  & 94 & 227 \\
Beaming  & 106 & 11 & 321 & 41 & 438 & 62 \\
\hline
\end{tabular}
\caption{Population of simulated radio-selected ($N_r$) 
and $\gamma$-ray-selected ($N_g$) MSPs for six-month, 
five-year and ten-year $Fermi$ observations. 
The first line; the results  for ten radio surveys 
listed in table~1 of Takata et al. (2011a). The second and third lines are 
 the results with  the sensitivities increased  by a factor of two and ten, 
respectively, and  the bottom is the populations associated with only
 beaming effects of the radio emission.  See Takata et al. 2011c.}
\end{table}

\begin{figure}
 \includegraphics[width=8cm, height=8cm]{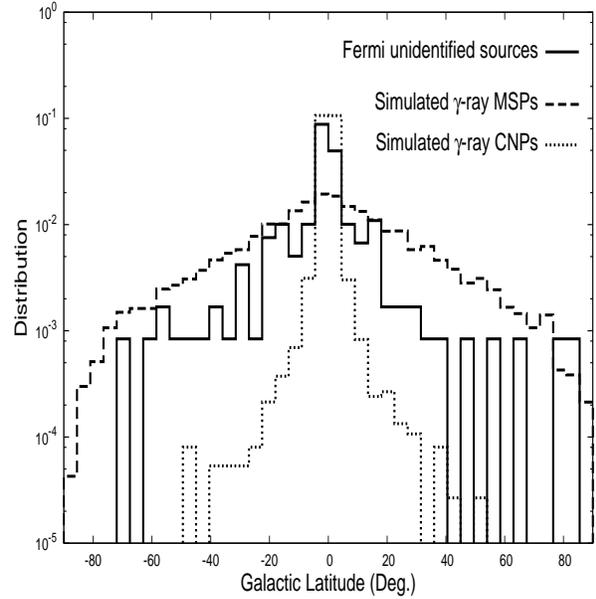}
 \caption{The distributions of  Galactic latitudes 
of the $Fermi$ unidentified sources with
 $V\le 23.21$ and $C \ge  5$ (solid line),
the simulated $\gamma$-ray MSPs (dashed line) and canonical 
pulsars (dotted line). See Takata et al. (2011b). }
\label{galbMSP}
 \end{figure}

\subsection{Population}
In Table~1, we summarize the simulated population of  MSPs 
 with the $Fermi$ six-month, five-year and ten-year long  observations. 
In addition, the second line (``Ra. Sen. (x2)'') and the third 
line (``Ra.Sen. (x10)'') in the tables show the results for the 
radio-surveys increased the sensitivities by a factor of two and of ten, 
respectively,  and
 the fourth line (``Beaming'') shows the population associated with 
the only beaming effect of the radio emission. 

 With  the previous radio surveys (first line in Table~1), the present 
simulation predicts  10 of the radio-selected 
$\gamma$-ray MSPs, which is consistent with the observed number $9$ of the $Fermi$ six-month long observation.  
The present model predicts that  14 (or 16) radio-selected $\gamma$-ray MSP 
pulsars will be detected by the $Fermi$ with five-year (or ten-year)  observations.
 
We see in the first line of Table~1  that  the simulated 
 numbers  of radio-selected $\gamma$-ray  MSPs increase
only $\sim$10 sources, respectively, over even  ten-year $Fermi$  
observations. This implies that most presently
known radio MSPs ($\sim 80$) might not be discovered by the  $Fermi$ future 
observations.   We also see in Table~1,  most  simulated $\gamma$-ray 
emitting  MSPs  are categorized as the $\gamma$-ray-selected pulsars 
with the previous sensitivities of the radio surveys, although 
the $Fermi$ has not confirmed the radio-quiet MSPs.  
We argue  that  it  may be difficult to identify  radio-quiet  
 MSPs, because   the detection of the rotation period 
by the $Fermi$  blind search is very difficult due to, for example, 
the effect orbital motion  if the MSP is in binary system. We expect 
that  the  $\gamma$-ray-selected MSPs in the simulation correspond
 to  the $Fermi$ unidentified sources located at higher Galactic latitudes. 
 
We would like to remark  that even we drastically increase the radio 
sensitivity by a factor of 10, the number of radio-selected MSPs 
 detected by 10~year $Fermi$ observations
 can increase from  16 to 94, which is still 
much less than the expected 227 $\gamma$-ray-selected MSPs.
 Unless the $Fermi$  sensitivity of the blind search is improved,  therefore, 
the most $\gamma$-ray MSPs will not be identified and will
 contribute to the $Fermi$ unidentified sources and/or the 
$\gamma$-ray background radiations. 

\subsection{$Fermi$ unidentified sources}
In Figure~\ref{galbMSP}, we compare 
the distributions of the Galactic  latitudes for the unidentified $Fermi$
sources with the variability index $V\le 23.21$ and curvature index  
$C\ge 5$ (solid line, Abdo et al. 2010b), 
the simulated $\gamma$-ray MSPs (dashed line), and canonical pulsars 
(dotted line)  with a flux $F_{\gamma}\ge\ 10^{-11}~\mathrm{erg/cm^2 s}$. 
 
We find in Figure~\ref{galbMSP} that the distribution of
the simulated $\gamma$-ray MSPs  is  consistent
 with that of the $Fermi$ unidentified sources. In particular, 
the MSPs can explain the distribution of the $Fermi$ unidentified sources 
located above the Galactic plane $|b|\ge 10$, which cannot be explained by the
canonical $\gamma$-ray  pulsars, which can mainly explain the
unidentified sources located in the Galactic plane.
Since the MSPs are in general older
than  the canonical  pulsars, a higher
fraction of the $\gamma$-ray MSPs, as compared with the
canonical pulsars,  is located at
  higher Galactic latitudes. Hence,   we expect 
 that $\gamma$-ray emitting MSPs are more plausible as 
 the candidate for the origin of
 the majority of the Galactic $Fermi$ unidentified
steady sources located in  high Galactic latitudes.

Finally, several MSPs in the 
 Black Window (B-W) systems have been discovered at the $Fermi$ unidentified 
sources (Ransom et al. 2011). Because the MSPs in the B-W systems 
are younger and have higher spin-down power, 
the  $\gamma$-ray luminosity will  be larger than
 that of the ordinary MSPs. Furthermore,  
  because the $\gamma$-ray emission from the outer gap is
 stronger in the direction perpendicular to the spin axis, 
 $Fermi$ is more likely to discover a greater number of MSPs 
with the Earth viewing angle $\sim 90^{\circ}$ measured 
from the rotation axis (Takata, Cheng \& Taam 2010a; 
Takata et al. 2011c). 
 If  the  angular momentum transferred from the accreting 
 matter to the neutron star  in the accreting 
stage produces the pulsar's spin axis  perpendicular to the orbital plane,
  the $\gamma$-ray emissions from MSPs in B-W system  will be
 stronger on the orbital plane. Hence, the $Fermi$ will find the 
B-W systems with  the Earth viewing angle described by   
 edge-on rather than by face-on with respect 
 to the orbital plane. In such a case,  a 
greater number of the  $Fermi$ Black Window systems show 
the eclipse of the radio emissions by the matter ejected 
from the companion star.

\vspace{1zw}
In summary, we have studied the population of the $\gamma$-ray emitting MSPs. 
We find that the observed possible relation
 between $L_{\gamma}$ vs. $L_{sd}$ can be 
explained by the outer gap model controlled by the
 magnetic pair-creation process. Our Monte-Carlo study implies 
that the $Fermi$  has detected a greater number of  MSPs with 
 larger inclination  and larger  viewing angles. 
Furthermore, we expect that  most of $\gamma$-ray emitting 
MSPs has been missed by the $Fermi$ observations, 
and still remain as the unidentified sources. It is likely that the Galactic 
$Fermi$ unidentified sources located 
high galactic latitude are mainly  associated with MSPs.

\vspace{1zw}
 We thank   A.H.Kong, C.Y.Hui,  P.H.T.Tam, R.H.H.Huang, 
 Lupin-C.C.Lin, M.Ruderman,  S.Shibata  and R.Taam 
for the useful discussions.  K.S.C. are supported by a GRF grant of
 Hong Kong Government  under HKU700911P.


\end{document}